# GIGJ: a crustal gravity model of the Guangdong Province for predicting the geoneutrino signal at the JUNO experiment

M. Reguzzoni[1,2,3], L. Rossi[1], M. Baldoncini[3,4], I. Callegari[5], P. Poli[6], D. Sampietro[2], V. Strati[4,7], F. Mantovani[4,7], G. Andronico[8], V. Antonelli[9], M. Bellato[10,11], E. Bernieri[13,12], A. Brigatti[9,14], R. Brugnera[10,11], A. Budano[13], M. Buscemi[8,15], S. Bussino[12,13], R. Caruso[8,15], D. Chiesa[16,17], D. Corti[11], F. Dal Corso[10,11], X. F. Ding[9,18], S. Dusini[10,11], A. Fabbri[12,13], G. Fiorentini[4,7], R. Ford[19,9], A. Formozov[9,14,20,21], G. Galet[10,11], A. Garfagnini[10,11], M. Giammarchi[9,14], A. Giaz[10,11], M. Grassi[22], A. Insolia[8,15], R. Isocrate[10,11], I. Lippi[10,11], F. Longhitano[8], D. Lo Presti[8,15], P. Lombardi[9,14], Y. Malyshkin[13], F. Marini[10,11], S. M. Mari[12,13], C. Martellini[12,13], E. Meroni[9,14], M. Mezzetto[11], L. Miramonti[9,14], S. Monforte[8], M. Montuschi[4,7], M. Nastasi[16,17], F. Ortica[23,24], A. Paoloni[25], S. Parmeggiano[9,14], D. Pedretti[26,6], N. Pelliccia[23,24], R. Pompilio[9,14], E. Previtali[16,17], G. Ranucci[9,14], A. C. Re[9,14], B. Ricci[4,7], A. Romani[23,24], P. Saggese[9,14], G. Salamanna[12,13], F. H. Sawy[10,11], G. Settanta[12,13], M. Sisti[16,17], C. Sirignano[10,11], M. Spinetti[25], L. Stanco[10,11], G. Verde[8], L. Votano[25]

[1]Politecnico di Milano, Department of Civil and Environmental Engineering DICA, Milan, Italy.

[2]Geomatics Research & Development GReD s.r.l., Lomazzo CO, Italy.

[3] INFN, Legnaro National Laboratories, Padua, Italy.

[4]University of Ferrara, Department of Physics and Earth Sciences, Ferrara, Italy.

[5]German University of Technology, Department of Applied Geosciences AGEO, Muscat, Oman.

[6]Université Grenoble-Alpes, Isterre Laboratory, Grenoble, France.

[7]INFN, Ferrara Section, Ferrara, Italy.

[8]INFN, Catania Section, Catania, Italy.

[9]INFN, Milan Section, Milan, Italy.

[10]University of Padua, Department of Physics and Astronomy, Padua, Italy.

[11]INFN, Padua Section, Padua, Italy.

[12]Roma Tre University, Department of Mathematics and Physics, Rome, Italy.

[13]INFN, Roma Tre Section, Rome, Italy.

[14]University of Milan, Department of Physics, Milan, Italy.

[15]University of Catania, Department of Physics and Astronomy, Catania, Italy.

[16]University of Milano-Bicocca, Department of Physics, Milan, Italy.

[17]INFN, Milano-Bicocca Section, Milan, Italy.

[18]Gran Sasso Science Institute, L'Aquila, Italy.

[19]SNOLAB, Lively, Ontario, Canada.

[20]Joint Institute for Nuclear Research, Dubna, Russia.

[21]Lomonosov Moscow State University, Skobeltsyn Institute of Nuclear Physics, Moscow, Russia.

[22]Astroparticule et Cosmologie, Paris, France.

[23]University of Perugia, Department of Chemistry, Biology and Biotechnology, Perugia, Italy.

[24]INFN, Perugia Section, Perugia, Italy.

[25]INFN, Frascati National Laboratories, Frascati, Italy.

[26]University of Padua, Department of Information Engineering, Padua, Italy.

Corresponding author: Mirko Reguzzoni (mirko.reguzzoni@polimi.it)

**Abstract**

Gravimetric methods are expected to play a decisive role in geophysical modeling of the regional crustal structure applied to geoneutrino studies.

GIGJ (GOCE Inversion for Geoneutrinos at JUNO) is a 3D numerical model constituted by ~46×10$^3$ voxels of 50 × 50 × 0.1 km, built by inverting gravimetric data over the 6° × 4° area centered at the Jiangmen Underground Neutrino Observatory (JUNO) experiment, currently under construction in the Guangdong Province (China). The a-priori modeling is based on the adoption of deep seismic sounding profiles, receiver functions, teleseismic P-wave velocity models and Moho depth maps, according to their own accuracy and spatial resolution. The inversion method allowed for integrating GOCE data with the a-priori information and regularization conditions through a Bayesian approach and a stochastic optimization. GIGJ fits the homogeneously distributed GOCE gravity data, characterized by high accuracy, with a ∼1 mGal standard deviation of the residuals, compatible with the observation accuracy.

Conversely to existing global models, GIGJ provides a site-specific subdivision of the crustal layers masses which uncertainties include estimation errors, associated to the gravimetric solution, and systematic uncertainties, related to the adoption of a fixed sedimentary layer. A consequence of this local rearrangement of the crustal layer thicknesses is a ~21% reduction and a ~24% increase of the middle and lower crust expected geoneutrino signal, respectively. Finally, the geophysical uncertainties of geoneutrino signals at JUNO produced by unitary uranium and thorium abundances distributed in the upper, middle and lower crust are reduced by 77%, 55% and 78%, respectively. The numerical model is available at http://www.fe.infn.it/u/radioactivity/GIGJ

# 1    Introduction

Understanding the composition of the Earth is a puzzling question that continuously pushes the scientific community to conceive innovative methods for gathering access to the interior of our planet.

While the geophysical structure of the entire Earth is almost well established, available information on its composition relies on shallow drill cores and samples brought to the surface by volcanic eruptions. Breakthroughs in the field are expected from the interplay between Earth Science and Particle Physics, which are currently exploring the promising scenarios of the Earth's spectrometry with atmospheric neutrino oscillations (Rott et al., 2015) and the detection of geoneutrinos (Fiorentini et al., 2007).

Geoneutrinos are electron antineutrinos produced in beta decays of naturally occurring radioactive isotopes in the Earth: they propagate almost without interacting, providing instantaneous insights on the radiogenic heat power of our planet. The present technology permits to detect geoneutrinos produced by beta decays of $^{234m}$Pa and $^{214}$Bi ($^{238}$U decay series) and $^{228}$Ac and $^{212}$Bi ($^{232}$Th decay series). By measuring their flux and energy spectrum it is possible to infer the global amount, distribution and ratio of U and Th in the crust and in the mantle, essential ingredients for the discrimination among different bulk silicate Earth compositional models (Šrámek et al., 2013). Recent measurements from the KamLAND (Japan) (Gando et al., 2013) and Borexino (Italy) (Agostini et al., 2015) experiments are opening the way to multiple-sites geoneutrino studies aimed at distinguishing the site-dependent crustal components (~75% of the signal) from the mantle component (~25% of the signal) (Fiorentini et al., 2012). In this framework, new geoneutrino measurements are highly awaited from the SNO+ detector (Canada) (Andringa et al., 2016) and from the Jiangmen Underground Neutrino Observatory (JUNO) experiment (An et al., 2016).

The JUNO experiment is under construction in Kaiping, Jiangmen, Guangdong Province (South China), 53 km far from two nuclear power plants, which is the optimum distance for the determination of the neutrino mass hierarchy from reactor antineutrino oscillation interferences. The 20 kton liquid scintillation detection volume, together with the excellent energy resolution, will allow JUNO to address many physics goals related to the observations of neutrino events of astrophysical and terrestrial origin (An et al., 2016; Strati et al., 2015).

Since the beginning of 1900, when the Croatian seismologist Andrija Mohorovičić discovered its existence, the study of the crust-mantle discontinuity (Moho) and more in general of the lithosphere architecture has been mainly performed by seismic observations. In 2012 the GEMMA project, funded by the Politecnico di Milano and the European Space Agency (ESA), demonstrated the possibility of exploiting satellite gravity data from the Gravity field and steady-state Ocean Circulation Explorer (GOCE) mission (Drinkwater et al., 2003) to model the main features of the crust at both global and regional scales (M. Reguzzoni & Sampietro, 2015).

The use of satellite data offers the main advantage of giving a regional outline of the crustal architecture that integrates data from local seismic profiles which are often not homogenously distributed. In this respect, satellite gravity observations, especially those coming from the GOCE mission, can be considered the optimal tool to study the main features of the Earth's crust at regional scale, for which resolutions better than 30-50 km are not really required. The major issues when using gravity observations for crustal modeling are the non-uniqueness and the ill-posedness of the problem. As it is well known, see e.g. (Sampietro & Sansò, 2012), the inverse gravimetric problem, i.e. the estimation of the masses generating a gravitational field from observations of the field itself, does not generally have a unique solution. Moreover, the problem is strongly unstable and requires some kind of regularization to control the effects of observation and model errors. Several approaches have been studied in literature to solve inverse gravimetric problems, see e.g. (Blakely, 1996; Parker, 1994) and the references therein.

To tackle this challenge we used a new algorithm, based on a Bayesian approach (Bosch, 2004; Mosegaard & Tarantola, 2002; Rossi et al., 2016), and able to invert the gravity field by also exploiting some a-priori information on the crustal structure derived from a combination of geological maps and seismic data. This helped us to reduce the ill-posedness and non-uniqueness of the problem, thus obtaining a 3D voxel-wise crustal model beneath the Guangdong province, south-eastern China, to be used for predicting the geoneutrino flux at JUNO. Following the approach described in (Coltorti et al., 2011; Huang et al., 2014) the 3D model, centered at the location of the JUNO detector, covers an area of 6° × 4° from which 50% of total geoneutrino signal is expected (Strati et al., 2015).

## 2   Geological setting of the region

The study area is located in the south-east of China and includes the northern margin of the South China Sea (SCS), the Guangdong region and the south-eastern part of Guangxi region (Figure 1). It is a part of the South China Block (SCB) that has a complex tectonic history (John et al., 1990; Zeng et al., 1997), as well as a composition and a thickness poorly understood (Zheng & Zhang, 2007). The SCB is composed of two collided Neoproterozoic continental crustal blocks (He et al., 2013): the Yangtze, in the north-west sector, which forms a stable cratonic area, and the Cathaysia Block (CB) in the south-east (Xu et al., 2007), which comprises the offshore continental margin of the SCS (Pearl River Mouth Basin); the boundary between these blocks is still object of debate (Deng et al., 2014; He et al., 2013; Xia et al., 2015).

The CB consists of Palaeo and Mesoproterozoic intensely folded basement rocks (gneisses, amphibolites and migmatites) with superimposed Mesozoic and early Cenozoic volcanism and granitic intrusions (a total area of ~220˙000 km$^2$), covered by Sinian to Mesozoic sedimentary and volcanic rocks. The granitoids of CB have interested various tectonic settings, with heterogeneous sources and repeated processes of crustal melting, mixing and fractional crystallization (Jiang et al., 2009; Wang et al., 2010).

Starting from the north-west to the south-east, the continental crust is characterized by lateral variations in thickness and composition, as well as in P-wave velocity, reaching a transition zone that continues until the oceanic crust of the SCS (Li et al., 2007). The crust exhibits a typical layer distribution into Upper Crust (UC), Middle Crust (MC), and Lower Crust (LC) and refers to a felsic and intermediate composition (Li et al., 2007), mainly for the continental sector. In the transition zone toward the oceanic crust, south-east of the 6°×4° area centered in JUNO (Figure 1), this distinction is less recognizable. The CB shows a general younger trend, from inland to coast, and an increasing occurrence of intrusions (mainly in the upper portion of the crust) moving from north-west to north-east.

The CB can be subdivided into three parts by two distinct tectonic regional elements (Figure 1), the Shi-Hang Zone (SHZ) and the Lishui-Haifeng Fault (LHF), which from north-west to south-east are: the Cathaysia Interior (CI), the Cathaysia Folded Belt (CFB) and the Southeast Coast Magmatic Belt (SCMB) (Chen et al., 2008; Xia et al., 2015). The SHZ has been interpreted as an intra-arc rift (back-arc extensional zone related to the paleo-Pacific plate subduction) that affected the middle to late Jurassic felsic and mafic magmatism in south-east China; it played an important role in the reworking of the crust and lithosphere in the study region (Jiang et al., 2009; Xia et al., 2015). Together with the high angle strike slip fault (LHF), this element appears to be a discriminating factor for the distribution of Mesozoic magmatic rocks. The Triassic granites are mainly distributed in the CI and CFB, the Cretaceous granitoids in the SCMB and the Jurassic rocks in the CFB (Chen et al., 2008). The SCMB consists of intermediate to mafic compositions, compared to the felsic compositions in the CFB and the CI (Xia et al., 2015). We emphasize that the CB is characterized by the exposition of widespread Mesozoic granitic and volcanic rocks, particularly in the coastal area, and by a slightly decreasing degree of acidity moving from west to east.

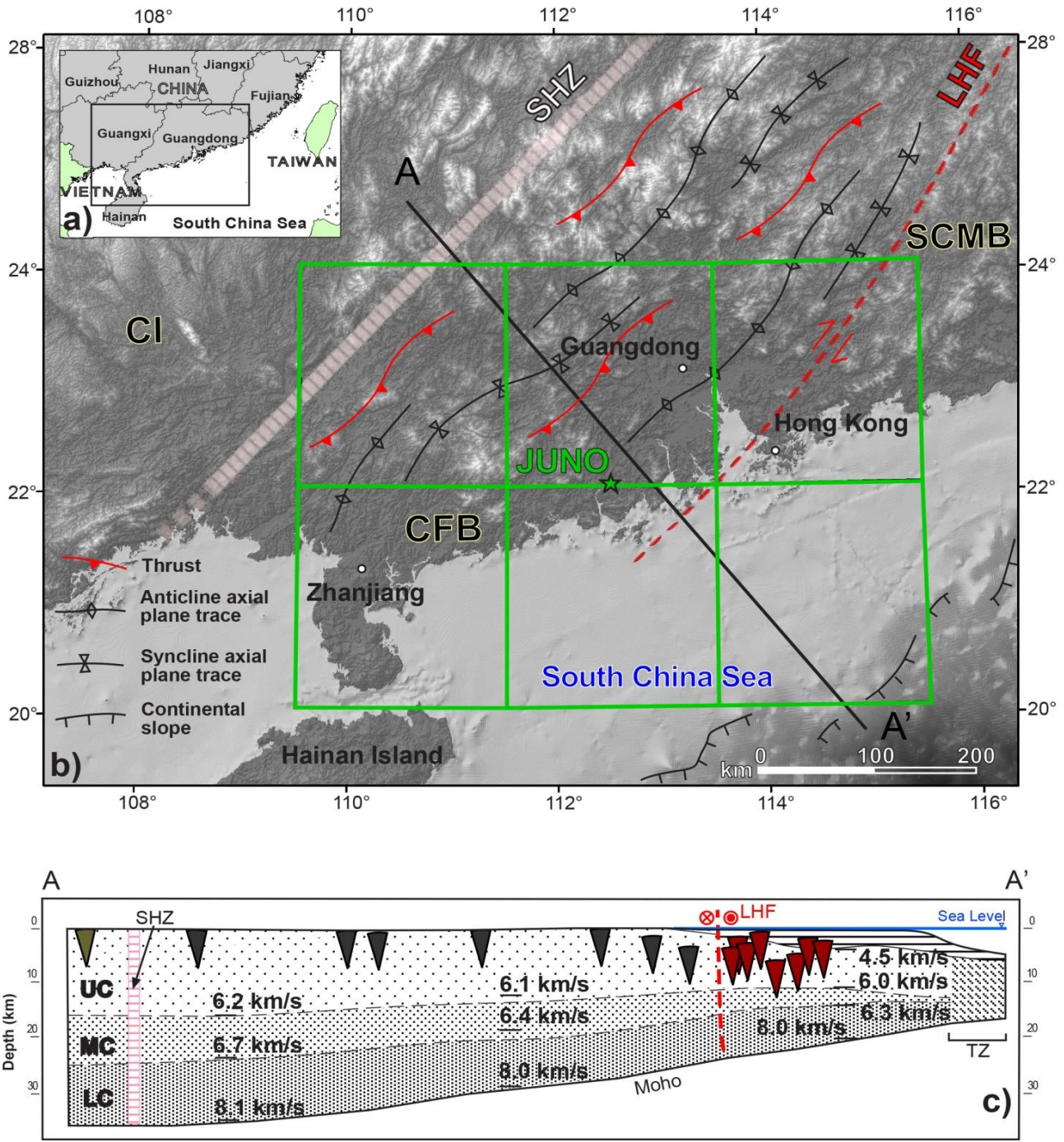

**Figure 1. a)** Location map of the study area outlined with a black rectangle. **b)** Simplified tectonic map of the region in which the JUNO detector will be located (green star): the tectonic partition of the Cathaysia Block is divided into three parts, the Cathaysia Interior (CI), the Cathaysia Fold Belt (CFB) and the South-east Coast Magmatic Belt (SCMB), on the basis of two regional geological features, the Shi-Hang Zone (SHZ) and the Lishui-Haifeng Fault (LHF), (modified after (Chen et al., 2008)). The 3D regional crustal model refers to the 6° × 4° area (depicted with green rectangles) centered at JUNO. **c)** Schematic crustal cross-section showing the vertical layer distribution (Upper Crust - UC, Middle Crust - MC and Lower Crust - LC) inferred from seismic data on the basis of average P-wave velocity. A lateral variation moving to the Transition Zone (TZ) towards the oceanic crust is visible. The top layer (parallel black lines) represents the sedimentary cover (Sinian-Mesozoic). The colored triangles highlight the increasing and typically younger trend of the intrusion from north-west to south-east; in grey the Triassic intrusions, in black the Jurassic ones and in red the Cretaceous ones. The LHF has been interpreted as a regional strike-slip fault with right movement and limited to the lower crust (modified after (Zhou et al., 2006)).

## 3   Geophysical datasets

As our final model assumes a layered crust and a one-layer uppermost mantle, we defined the following surfaces: the topography/bathymetry, the bottom of sediments, the Top of the Upper Crust (TUC), the Top of the Middle Crust (TMC), the Top of the Lower Crust (TLC), the Moho Discontinuity (MD) and a horizon with a constant depth of 50 km, which is the bottom of the model.

Constraints for the definition of the crustal model were obtained from published studies including Deep Seismic Sounding profiles (DSS), Receiver Functions (RF), teleseismic P-wave velocity models and Moho depth maps (Figure 2). The inputs and their corresponding uncertainty are summarized in Table A1 and A2 of the Appendix, which details the criteria used for the selection, interpretation and implementation of the data in the a priori model.

The P-wave velocities for the different crustal layers were obtained from DSS belonging to 12 seismic profiles (Figure 2). The P-wave velocity contours of each model were used as benchmarks for the depth of the modeled geophysical surfaces. The $3\sigma$ MD uncertainty associated to each DSS profiles was estimated by considering the picking error ($1\sigma$) from each reference paper (Table S2). The uncertainties for the TLC and TMC were subdivided into two quality classes based on the clarity of the corresponding velocity contour during the digitalization (Table S2).

Additional punctual constraints for the MD come from 10 teleseismic stations, 2 located inside and 8 outside the study area (Figure 2). As the MD information from the stations are provided according to different analysis methods (Tkalčić et al., 2011), the $3\sigma$ MD uncertainty was estimated by accounting both for the individual uncertainty of each method and for the variability of the different MD data (Table S2).

We further used a 3D P-wave velocity model from (Sun & Toksöz, 2006). The $3\sigma$ MD uncertainty was estimated on the basis of the standard deviation of the travel time of the final model and on the basis of the mean velocity in the lower crust.

Finally, 3 Moho depth maps were adapted from (Hao et al., 2014; He et al., 2013; Xia et al., 2015) and provided in the construction of the a priori model as regular grids with 10 km × 10 km horizontal resolution.

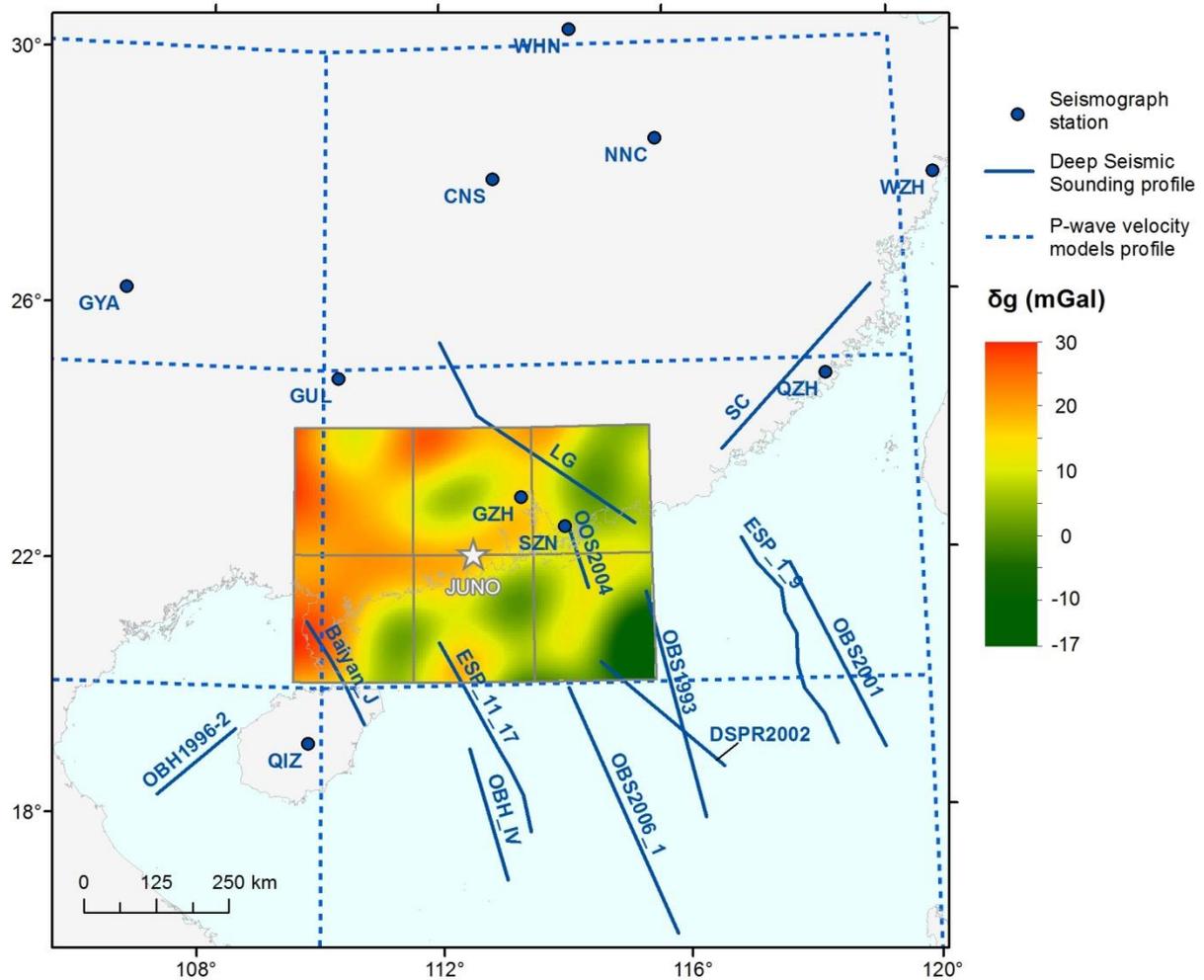

**Figure 2**. Geophysical input data used for the construction of the 3D model in the 6° × 4° area centered at the JUNO detector location (22° 07' 05" N, 112° 31' 05" E) (An et al., 2016). Deep seismic sounding profiles, P-wave velocity profiles and locations of seismograph stations correspond to the input seismic data used to build the a priori model for the inversion of gravimetric data. The raw observations of gravimetric disturbances (δg) are represented as a continuous grid with 5 km × 5 km resolution.

# 4 Bayesian gravity inversion

In the next sections, the method of the gravity inversion in the JUNO area will be shown. The Bayesian mathematical model of the problem is presented, including some additional constraints that can be put on the solution, together with the gravity data and the a-priori information that are provided as input to the mathematical model.

## 4.1 Target function of the gravity inversion problem

The investigated volume is split into voxels, $V_i$, with index $i = 1, 2, ..., N$. Each voxel is a regular prism with a fixed size and it is described by two parameters: a label $L_i$, denoting the material constituting the voxel (e.g. UC, MC, etc.), and a mass density $\rho_i$, that is assumed constant inside the voxel volume. The prisms were disposed on a regular grid in Cartesian coordinates, and the forward modeling was performed in planar approximation. This geometry allowed easily introducing neighborhood relationships that will be discussed later.

The inversion algorithm is based on the Bayes theorem:

$$P(\boldsymbol{x}|\boldsymbol{y}) \propto \mathcal{L}(\boldsymbol{y}|\boldsymbol{x})P(\boldsymbol{x})$$

(1)

where $\boldsymbol{y}$ is the vector of observables, i.e. the gravity signal, $\boldsymbol{x}$ is the vector of parameters $\boldsymbol{L}, \boldsymbol{\rho}$ for all voxels, $P(\boldsymbol{x}|\boldsymbol{y})$ and $P(\boldsymbol{x})$ are the posterior and the prior distribution, respectively, and $\mathcal{L}(\boldsymbol{y}|\boldsymbol{x})$ is the likelihood. Since gravity is observed, the likelihood represents the degree of fit between the observed signal $\boldsymbol{y}_o$ and the modelled one $\boldsymbol{y}(\boldsymbol{x})$. Its distribution was derived from the assumption that the observation noise $\boldsymbol{\nu}$ is normal, namely:

$$\mathcal{L}(\boldsymbol{y} = \boldsymbol{y}_o|\boldsymbol{x}) \propto \exp\left\{-\frac{1}{2}[\boldsymbol{y}_o - \boldsymbol{y}(\boldsymbol{x})]^T \mathbf{C}_{\nu\nu}^{-1}[\boldsymbol{y}_o - \boldsymbol{y}(\boldsymbol{x})]\right\}$$

(2)

with $\mathbf{C}_{\nu\nu}$ the noise covariance matrix. It is worth to notice that, even if theoretically $\boldsymbol{y}(\boldsymbol{x})$ depends on the full set of parameters, in this case it depends directly on the full set of densities $\boldsymbol{\rho}$ only. The other parameters, i.e. the labels $\boldsymbol{L}$, act indirectly through the prior distribution.

The prior distribution was defined by considering the available geophysical information on the study region, integrated with some regularization conditions. This information was supplied to the algorithm in the following way:
- a range of variation of each boundary surface between two layers with different labels;
- neighborhood rules between the possible couple of labels;
- the density of each material, i.e. of each label, in terms of the most probable value and its range of variation.

The shape of the prior distribution was chosen to highlight the dependency of each density $\rho_i$ on the label $L_i$:

$$P(\mathbf{x}) = \prod_{i=1}^{N} P(\rho_i|L_i) \cdot P(\mathbf{L})$$

(3)

The density of each voxel $\rho_i$ was assumed to be normally distributed once the label $L_i$ was given:

$$P(\rho_i|L_i) \propto \exp\left\{-\frac{\left(\rho - \mu_\rho(L_i)\right)^2}{2\sigma_\rho^2(L_i)}\right\}$$

(4)

where the mean $\mu_\rho(L_i)$ and the variance $\sigma_\rho^2(L_i)$ were given as a-priori information.

On the other hand, the labels $\mathbf{L}$ were modeled as a Markov Random Field. Therefore, their probability distribution assumed the shape of a Gibbs distribution, where the energy is the sum of the clique potential (Azencott, 1988):

$$P(\mathbf{L}) \propto \exp\left\{-\frac{1}{2}\gamma \sum_{i=1}^{N} s_i^2(L_i) - \frac{1}{2}\lambda \sum_{i=1}^{N} \sum_{j \in \Delta_i} q^2(L_i, L_j)\right\}$$

(5)

where $s_i^2(L_i)$ and $q^2(L_i, L_j)$ are two penalty functions defining and weighting the admissible labels for each voxel $i$ and for its neighbour $\Delta_i$, respectively, and $\gamma$ and $\lambda$ are the relative weights between these two penalty functions. The most probable label realization is the one with the smallest overall penalty.

The function $s_i^2(L_i)$ is used to define the limits of the boundary surfaces between two layers. In particular, it is equal to 0 if the label $L_i$ is admissible for the voxel $i$, otherwise it is equal to $+\infty$, i.e. the highest penalty value. This setup implies that the value of the weight $\gamma$ is irrelevant.

The function $q^2(L_i, L_j)$ is used to define neighbourhood rules between different materials, thus controlling the smoothness of the boundary surfaces and preventing layers with null thickness. In particular, it is equal to 0 if the neighbors labels $L_i$ and $L_j$ are the same, equal to 1 if they are different but their closeness is admissible, and equal to $+\infty$ if they are different and they cannot even be one close to the other. According to this definition, the higher is the value of the weight $\lambda$, the higher is the penalty for different neighbor labels, and therefore the higher is the smoothness imposed to the boundary surfaces.

Combining Eqs. (2), (3), (4) and (5) into Eq. (1), the posterior distribution is derived. Then, invoking the Maximum A Posteriori (MAP) principle, the most probable set of labels and densities were chosen as the solution. This corresponds to finding the minimum of the following target function:

$$F(\boldsymbol{\rho}, \boldsymbol{L}|\boldsymbol{y}_o) = [\boldsymbol{y}_o - \boldsymbol{y}(\boldsymbol{x})]^T \boldsymbol{C}_{vv}^{-1}[\boldsymbol{y}_o - \boldsymbol{y}(\boldsymbol{x})] + \eta \sum_{i=1}^{N} \frac{\left(\rho - \mu_\rho(L_i)\right)^2}{\sigma_\rho^2(L_i)} + \gamma \sum_{i=1}^{N} s_i^2(L_i)$$
$$+ \lambda \sum_{i=1}^{N} \sum_{j \in \Delta_i} q^2(L_i, L_j)$$

(6)

where the additional weight $\eta$ is equal to the ratio between the number of observations and the number of voxels. Its introduction into the target function is to balance the magnitude of the contributions due to the gravity residuals and the density variations from the mean. The minimum was retrieved by using a stochastic optimization method, i.e. a simulated annealing aided by a Gibbs Sampler (Robert & Casella, 2004).

### 4.2    Constraints on the solution domain

The minimization of Eq. (6) can lead to a solution of the inverse gravimetric problem that is optimal from the mathematical point of view, but with a questionable physical meaning. The idea to overcome this drawback is to restrict the domain of the acceptable densities, so as to guarantee the plausibility of the solution. In addition, this restriction has the benefit of numerically stabilizing the minimization process.

The first weak point of the mathematical model defined in Section 4.1 is that the prior distribution of the voxel density given the label is normal, see Eq. (4). This is an advantage for the computation of the posterior distribution but does not correspond to the reality, since generally each material has a finite density range (Telford et al. 1990). Although very unlikely, some unphysical density values may be attributed to a subset of voxels in order to minimize Eq. (6). The proposed solution is to choose $[\mu_\rho(L_i) - 3\sigma_\rho(L_i), \mu_\rho(L_i) + 3\sigma_\rho(L_i)]$ as the admissible density range for the label $L_i$. The solution of Eq. (6) is then searched into the hyper-parallelogram defined by the Cartesian product of the $6\sigma_\rho(L_i)$ density ranges of all the voxels $i$. The sides of this hyper-parallelogram can be reduced by introducing a scaling factor $\alpha_\rho$ ($0 \leq \alpha_\rho \leq 1$) of the density standard deviations $\sigma_\rho(L_i)$. This further restriction is useful to reduce the density variability inside each layer of the solution of the inverse gravimetric problem as gravity data could be equally well fitted by a concentrated or a disperse density model. The former is here preferred and its selection is obtained by increasing the value of $\alpha_\rho$.

Another weakness of the mathematical model defined in Section 4.1 is that, given the labels, the voxel densities are independent to one another, see Eq. (3). This means that a rough density model with sharp variations between close voxels is a very likely solution of the inverse gravimetric model, because the gravity fitting is reached by freely adapting the densities and maintaining very smooth boundary surfaces between layers. To avoid this result, the simplest approach would be the introduction of a non-diagonal density covariance matrix $\mathbf{C}_{\rho\rho}$ into Eq. (3). However, this choice would have severe computational implications in the stochastic minimization of the resulting target function. For this reason, the solution domain is restricted by adding some constraints on the lateral and vertical variation of the density inside each layer, which is an alternative (deterministic) way of introducing a density spatial correlation. The maximum admissible values of the lateral and vertical density variations are respectively defined by two scale factors $\alpha_\ell$ and $\alpha_v$ ($0 \leq \alpha_\ell \leq 1$, $0 \leq \alpha_v \leq 1$) of the rescaled density ranges, which are equal to $\alpha_\rho \cdot 6\sigma_\rho(L_i)$. A further restriction of the solution domain is applied to force a density increasing (UC, MC and LC) or decreasing (uppermost mantle layer) trend with depth inside each layer, which is well justified from the geological point of view.

As the minimization procedure is based on a Gibbs Sampler, the presented constraints are not simultaneously applied to the joint density distribution of all the voxels, but are sequentially applied to the conditional density distribution of each voxel given the others, making the evaluation of the solution domain much easier.

Finally, it should be noted that the normalization constant of the posterior distribution does depend on the shape of the density domain, and consequently on the parameters $\alpha_\rho$, $\alpha_\ell$ and $\alpha_v$. Since this normalization constant has not a simple analytical expression and its numerical evaluation would significantly increase the computational burden of the whole minimization procedure, it is not included into the target function of Eq. (6). This implies that the target function cannot be used to compare solutions based on different values of the parameters $\alpha_\rho$, $\alpha_\ell$ and $\alpha_v$. This is also the reason why these parameters are a-priori fixed and are not considered as random variables (hyper-parameters) with their own prior distributions.

### 4.3 Gravity data

The voxel model to be estimated was chosen with a horizontal resolution of 50 km × 50 km and a vertical one of 100 m. The horizontal resolution was designed according to the requirements for the geoneutrino flux computation, while the vertical one was chosen as a trade-off between gravity sensibility and expected variability of the sediments boundary surfaces.

Given these geometrical parameters, the observations to be inverted were the gravity anomalies synthesized from a global gravity model on a grid of 50 km horizontal resolution, namely the same of the voxel model, at an ellipsoidal height of 600 m, guaranteeing to be as closest as possible to the topography but outside masses. The global gravity model was chosen between a satellite-only solution, with the advantage of being computed by homogeneous data, and a combined solution, having a higher spectral resolution. In particular, the latest release of the GOCE-only space-wise model up to degree and order 330 (Gatti & Reguzzoni, 2017; Mirko Reguzzoni & Tselfes, 2009) and the combined EIGEN-6C4 model up to degree and order 2190 (Förste et al., 2014; Shako et al., 2014) were considered. The choice was performed by comparing the empirical auto-correlation function of the gravity disturbances synthetized from these two models (up to different degrees and orders) with the one of the forward signal of the a-priori most probable model of the region. To avoid the introduction of useless high frequencies that cannot be interpreted by a voxel model with the given geometrical resolution, the observed and a-priori signal should have a similar stochastic behavior. Figure 3 shows that the correlation length of the GOCE-only space-wise model truncated at degree and order 200 was very similar to the one of the signal generated by the a-priori model, therefore this was the chosen model for the observation synthesis. The corresponding commission error was of the order of 1 mGal ($10^{-5}$ m/s$^2$) for the whole study area, considering a diagonal noise covariance matrix in Eq. (2).

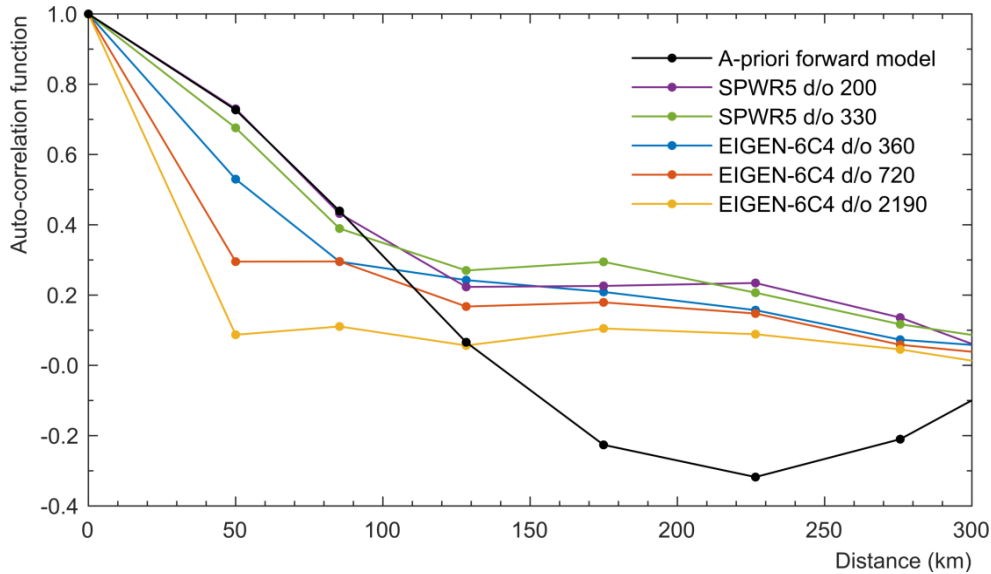

**Figure 3**. Empirical auto-correlation functions of the gravity disturbances synthetized from two global models truncated at different degrees and orders (d/o), compared to the auto-correlation function of the forward signal of the a-priori model. The correlation length was used as the figure of merit to choose the model adopted for the generation of the signal to be inverted.

**4.4 Prior crustal model**

The geophysical inputs shown in Section 3 are used to define the prior distribution in Eq. (3), providing geometrical and density information. Note that all the voxels above the TUC surface have fixed label and density because topography and bathymetry were taken from the GEBCO08 1 minute grid (Monahan, 2008), while the sedimentary layers were taken from the CRUST 1.0 model (Laske et al., 2013).

The a-priori geometrical information entered as the admissible depth ranges of the TMC, TLC and Moho surfaces. These ranges were computed by using the available geophysical data and their uncertainties (Figure 2). In the areas lacking in local seismic information, an additional input was given by the Refined Earth Reference Model (Huang et al., 2013), hereafter RRM, which provides the thickness with the corresponding uncertainty of the crustal layers at 1° × 1° spatial resolution. The depth data were firstly interpolated on the knots of the planimetric grid defined by the voxel model and then a range of 2 times the 3σ uncertainties (taken from Table S2) was opened around each depth value. Since there were many data sources, i.e. DSS, RF, teleseismic, depth maps and the RRM global model, the computation of a unique depth range for each knot and for each surface was required. This was performed by firstly joining depth ranges from the same sources and then intersecting the resulting ranges from different sources. The output was the admissible depth range of each boundary that is used to set up the penalty functions $s_i^2(L_i)$, see Eq. (5).

Concerning the a-priori density information to be used in Eq. (4), the mean and standard deviation of UC, MC and LC were defined as (2660 ± 80) kg/m$^3$, (2820 ± 20) kg/m$^3$ and (2980 ± 60) kg/m$^3$, respectively. These density values reproduce the statistics of the CRUST 1.0 global values, integrated with local values inferred from DSS seismic velocity data and the relationship between P-wave velocity and density given by (Christensen & Mooney, 1995). The uppermost mantle layer is a portion of the continental lithospheric mantle designed with a mean density decreasing in depth according to the PREM model (Dziewonski & Anderson, 1981) and a standard deviation of 100 kg/m$^3$.

Finally, an initial model was determined as a starting point for the simulated annealing. The boundary surfaces of this initial model were estimated through a regularized least squares adjustment of the geophysical inputs, disregarding the contribution of the gravity observations. These surfaces laid into the previously defined depth ranges and were as smooth as possible. The layer densities of each layer were fixed to the previously defined mean values. Since this initial model is seismic driven and gravity independent, it can be seen as a prior crustal model and can be also used to quantify the improvement brought by the gravity information to the final solution through the application of the Bayesian inversion, as discussed in Section 5.3.

Note that the gravity data cover an area of 6° × 4° as well as the final solution, but the inversion is actually performed on an area larger by a border of 3°. In this border, the initial model has the RRM boundary surfaces, properly adjusted to the available seismic profiles (see Figure 2), and the same homogeneous layer densities of the 6° × 4° area. Moreover, the forward modeling required for the computation of the likelihood (Eq. (2)) is based on an enlarged crustal model by a further border of 3°. This border is fixed to the RRM geometry and density distribution, with the aim of linking the inversion solution to a realistic, although approximate, crustal model. All these precautions in extending the working area has the main goal of making the final model much more robust against border effects, especially because of its small size of 6° × 4°.

## 5  The GIGJ model

In the following sections the selection criteria for choosing the best gravimetric solution are presented together with the output GIGJ model (GOCE Inversion for Geoneutrinos at JUNO) and geometry and density uncertainties of the crustal structure. The GIGJ model is further compared with the prior model and existing global crustal models.

### 5.1  Finding the best gravimetric solution

The final model was estimated by minimizing the target function of Eq. (6) for different sets of input parameters. In particular, the geometry smoothness was controlled by the value of the weight $\lambda$, the only free parameter of Eq. (6), while the density variability and smoothness were controlled by the parameters $\alpha_\rho$, $\alpha_\ell$ and $\alpha_v$ through the solution domain restriction (see Section 4.2). The considered values were the following:

- $\lambda = \bar{\lambda}$, 10 $\bar{\lambda}$, 100 $\bar{\lambda}$, 1000 $\bar{\lambda}$, 10000 $\bar{\lambda}$;
- $\alpha_\rho = 0.05, 0.10, 0.15, 0.20, 0.25, 0.50, 0.75, 1.00$;
- $\alpha_\ell = 0.05, 0.10, 0.15, 0.20, 0.50$;
- $\alpha_v = 0.05, 0.10, 0.15, 0.20, 0.50$;

where $\bar{\lambda} = 4 \times 10^{-3}$ is empirically computed from the prior model. Overall, this led to $10^3$ possible combinations of the input parameters for which the solution has to be retrieved. The computational burden of each solution was about 30 minutes on a standard personal computer, translating into about three days of computation by simultaneously running multiple processes on different machines.

An assessment of the different estimated models was performed to choose the best solution. A direct comparison of the target function values (Eq. (6)) was not useful, because of the missing normalization. Therefore, four indexes per each estimated solution were computed, which are $\sigma_g$, $r_l$, $r_v$ and $m$.

The $\sigma_g$ index evaluates the quality of the gravity fitting as the standard deviation of the residuals:

$$\sigma_g = \sqrt{\frac{1}{N_o}[\mathbf{y}_o - \mathbf{y}(\boldsymbol{\rho})]^T[\mathbf{y}_o - \mathbf{y}(\boldsymbol{\rho})]}$$

(7)

where $N_o$ is the number of observations.

Concerning the density smoothness, two quality indexes, $r_l$ and $r_v$, were introduced to separately evaluate the lateral and vertical density variations respectively. They were computed as the RMS of the maximum density differences between a voxel and its (horizontal or vertical) neighbors, namely:

$$r_k = \sqrt{\frac{1}{N}\sum_{i=1}^{N}\left(\max_{j \in \Delta_i^k}|\rho_i - \rho_j|\right)^2} \quad k = \ell, v$$

(8)

where $N$ is the number of voxels and $\Delta_i^k$ is the horizontal ($k = \ell$) or vertical ($k = v$) neighborhood of the voxel $i$.

As for the geometry smoothness, the estimated models were firstly translated in terms of discontinuity surfaces (i.e. TMC, TLC and MD) and then the quality index $m$ was computed as the RMS of the maximum slopes between a voxel and its neighbors, namely:

$$m = \sqrt{\frac{1}{3n} \sum_{j=1}^{3} \sum_{i=1}^{n} \left( \max_{j \in \Delta_i} \left| \frac{z_i^h - z_j^h}{d_{ij}} \right| \right)^2}$$

(9)

where $n$ is the number of knot of each surface, $\Delta_i$ is the neighborhood of each voxel $i$, $z^h$ is the depth of the surface $h$ (1 = TMC, 2 = TLC, 3 = MD), and $d_{ij}$ is the horizontal distance between the voxels $i$ and $j$.

The best solution was chosen according to a two-step procedure. Firstly, the solutions were filtered by imposing the following selection criteria:

- standard deviation of the gravity residuals $\sigma_g$ inside the range (1.0 ± 0.2) mGal to be compatible with the observation accuracy;
- RMS of the maximum slopes $m$ smaller than 2%; this threshold was chosen by rounding up the value of this index computed for the prior model used as starting point of the simulated annealing, which is equal to 1.89%.

The idea behind these two criteria is to look for a solution that is almost as smooth as the prior one but, differently from it, is able to fit the gravity observations. No constraints are put on the RMS of the maximum lateral and vertical density differences $r_l$ and $r_v$. Note that only an upper bound to $\sigma_g$ would have been strictly required to guarantee a proper gravity fitting. However, in order to obtain smaller gravity residuals, the algorithm would have produced solutions with stronger density variations among neighbor voxels, thus requiring the introduction of constraints on $r_l$ and $r_v$. Since these constraints would have been difficult to calibrate, this situation is avoided by setting a lower bound to $\sigma_g$.

The filtering procedure was passed by 17 solutions only, for which the values of the four indexes are shown in Figure 4. All the 17 solutions were fully consistent with the gravity observation accuracy in terms of $\sigma_g$. The best solution is then selected as the one for which the vector of the three indexes $r_l$, $r_v$, and $m$ has the minimum norm, after a proper normalization, namely the number 16 of the filtered set (Figure 4) corresponding to $\lambda = 1000\ \overline{\lambda}$, $\alpha_\rho = 0.2$, $\alpha_\ell = 0.2$ and $\alpha_v = 0.05$. Indeed, each of the three indexes is minimized by this solution, making their normalization irrelevant.

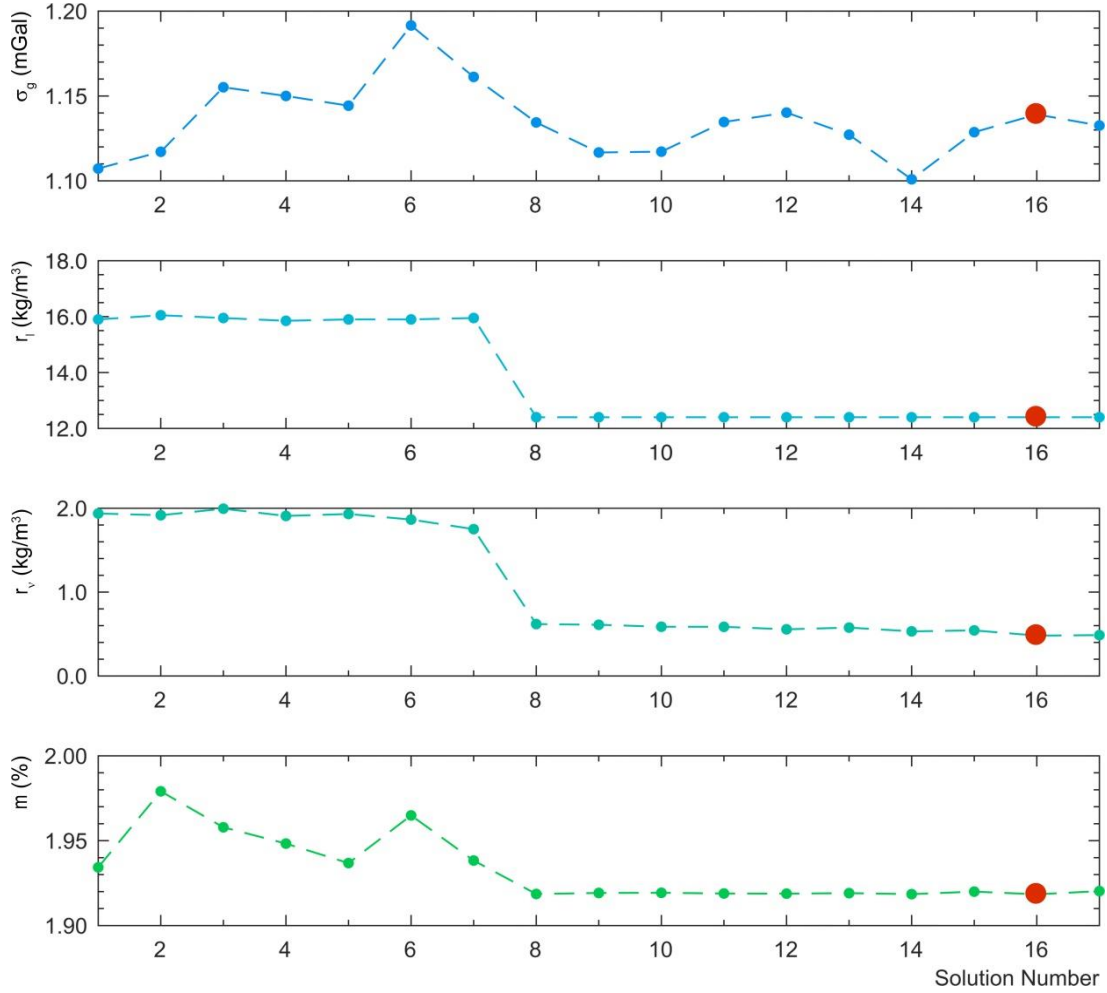

**Figure 4.** Values of the quality indexes defined in Eqs. (7), (8) and (9) for the filtered solutions, which were 17 out of 1000.

From now on, the selected model is called GIGJ (GOCE Inversion for Geoneutrinos at JUNO). Its gravity fitting is displayed in Figure 5, while its geometry and density distributions are shown in Figure 6 and Figure 7, respectively. As expected, the GIGJ crustal model exhibits a thinning of the crust moving from the continental area towards the oceanic region (i.e. along the north-west to south-east direction) (Figure 6), together with a higher spatial variability of the UC density with respect to the MC and LC layers (Figure 7).

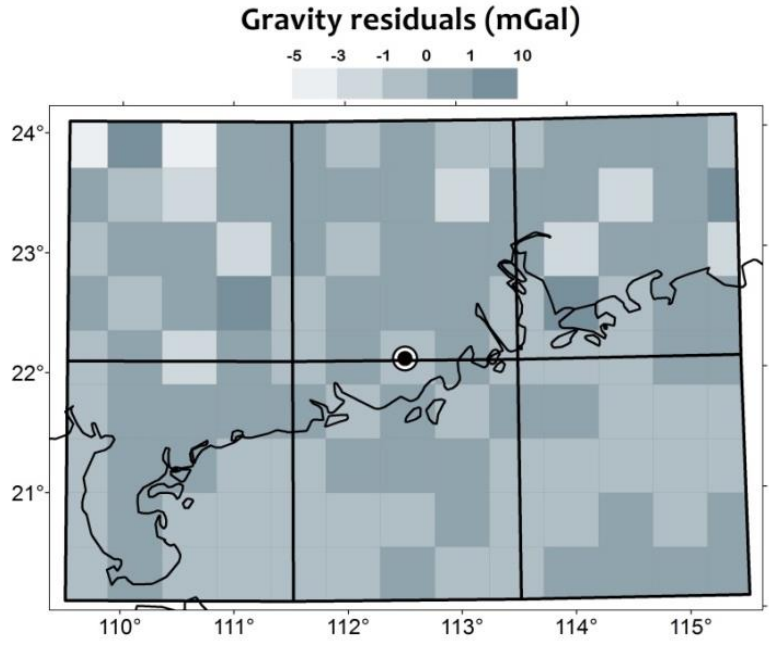

**Figure 5**. Gravity residuals of the GIGJ model.

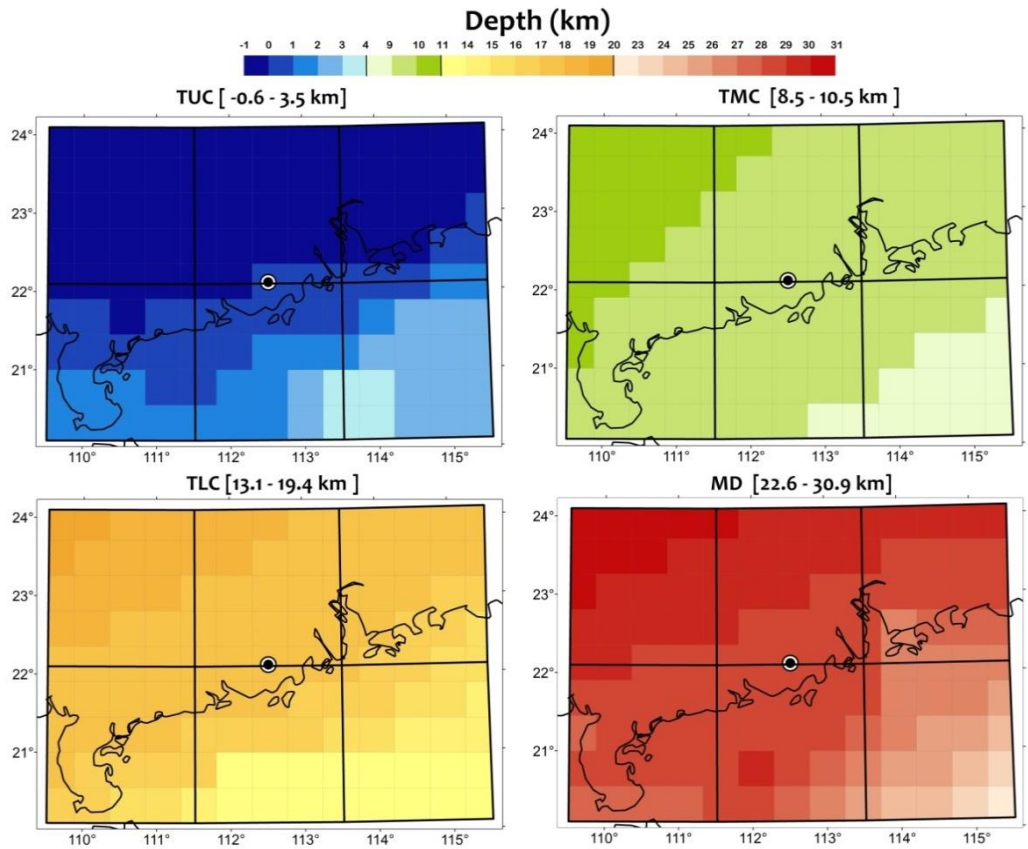

**Figure 6**. Depth maps of the Top of the Upper Crust (TUC), Top of the Middle Crust (TMC), Top of the Lower Crust (TLC) and Moho Discontinuity (MD) for the 6°×4° area centered at the JUNO detector location. Negative values mean surfaces above the zero-level.

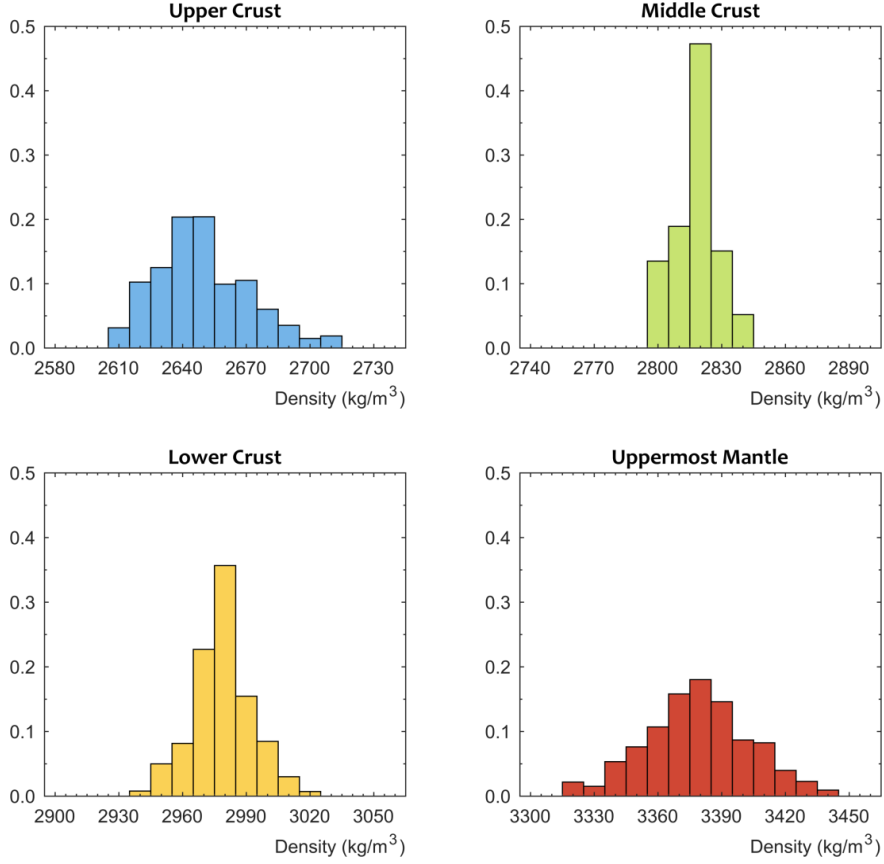

**Figure 7**. Frequency distributions of the density values for each label of the GIGJ model for the Upper Crust, Middle Crust, Lower Crust and Uppermost Mantle. The latter corresponds to the portion of continental lithospheric mantle down to 50 km of depth.

## 5.2 Model uncertainties

With the aim of estimating the geophysical contribution to the geoneutrino signal uncertainty for each crustal layer, the overall mass uncertainty was calculated for the GIGJ solution. It comprises an estimation error component associated to the solution of the inverse gravimetric problem and a systematic error component due to the adoption of a fixed sedimentary layer (Table 1).

The output of the GIGJ solution is made up of ~46×10$^3$ voxels, each one assigned with density and label values. The joint posterior distribution $P(\boldsymbol{\rho}, \boldsymbol{L} | \boldsymbol{y}_o)$ of all the voxels cannot be evaluated, neither analytically nor numerically. Therefore, the estimation error component of GIGJ was split into a density and a geometry contribution, both estimated by sample statistics on proper marginal distributions of the individual voxels; the sampling procedure was performed by using a Gibbs sampler, starting from the GIGJ solution, keeping fixed the corresponding values of $\lambda$, $\alpha_\rho$, $\alpha_\ell$ and $\alpha_v$ and drawing about 3×10$^6$ samples.

First, the density variability of each voxel $i$ was evaluated by sampling the marginal distribution $P(\rho_i | \boldsymbol{L}_{\text{GIGJ}}, \boldsymbol{y}_o)$ of the conditional posterior given the GIGJ label realization $\boldsymbol{L}_{\text{GIGJ}}$. The sample density variances were computed for all the voxels, and then averaged for each label (first column of Table 1).

Second, the geometrical variability of each voxel $i$ was evaluated by sampling the marginal posterior distribution $P(L_i|\mathbf{y_o})$. The probability that each voxel has a label different from the one of the GIGJ model was computed and then translated into a "volumetric uncertainty". This translation was performed by multiplying this probability by the voxel volume and summing over voxels belonging to the same label (second column of Table 1).

The reliability of the GIGJ model also depends on the accuracy of the sedimentary mass, which was a fixed input ($M_{SED} = 47.1 \times 10^{16}$ kg) in the a-priori model and which is known to be affected by a typical 15% relative uncertainty (Kaban & Mooney, 2001). By increasing (decreasing) the thickness of the sedimentary layers by 15% and by reapplying the inversion algorithm to the modified input data, the corresponding mass decrease (increase) of each crustal layer was computed in order to evaluate the systematic component in the mass uncertainty (Table 1). On the other hand, a density variation on the order of 15% produces a rearrangement of the mass distribution with a negligible overall mass variation of each crustal layer.

The MC presents a higher mass estimation error (2.5%) with respect to the UC (0.3%) and LC (0.7%), as the TUC is assumed to be known and the MD is relatively well marked by seismic information. On the contrary, the systematic component of the mass uncertainty is dominant for the UC (1.4%) and progressively decreases by respectively one and two orders of magnitude for the MC (0.3%) and the LC (0.03%). Finally, the $692.0 \times 10^4$ km$^3$ volume of uppermost mantle layer has a mean density of 3378.7 kg/m$^3$, where the corresponding low estimation error of 18.6 kg/m$^3$ results from its fixed bottom horizon.

**Table 1**. Mean density and volume of the Upper Crust (UC), Middle Crust (MC) and Lower Crust (LC) obtained for the GIGJ model[1].

|    | Density (kg/m$^3$) | Volume (10$^4$ km$^3$) | Mass (10$^{16}$ kg) |
| --- | --- | --- | --- |
| UC | 2649.3 ± 7.4 | 263.4 ± 0.1 | 697.8 ± 2.2 (± 9.6) |
| MC | 2818.0 ± 5.9 | 207.6 ± 5.2 | 585.0 ± 15.8 (± 1.9) |
| LC | 2978.7 ± 10.8 | 336.1 ± 1.9 | 1001.1 ± 9.3 (± 0.3) |

[1]The mean density and volume values are reported together with their estimation errors obtained by sample statistics on the marginal posterior distributions of the individual voxels. The relative mass estimation errors for each crustal layer are conservatively calculated summing the relative errors of density and volume, while the systematic uncertainties (in brackets) are obtained by evaluating the impact of a 15% sediment thickness variation.

It is important to underline that the estimated uncertainties in Table 1 do not account for variations of the parameters $\lambda$, $\alpha_\rho$, $\alpha_\ell$ and $\alpha_v$, that are fixed to the values of the optimal solution. However, by looking at Figure 4, one can easily realize that all the solutions from 8 to 17 could be considered almost equally smooth according to the three defined indexes, and therefore each of them could be selected as an alternative GIGJ model. This means that they can be used to assess the mass uncertainty depending on the choice of the parameters $\lambda$, $\alpha_\rho$, $\alpha_\ell$ and $\alpha_v$. The UC, MC and LC mass standard deviations of these ten solutions is 3.6 10$^{16}$ kg, 11.0 10$^{16}$ kg and 4.5 10$^{16}$ kg, respectively. Comparing these values with those reported in Table 1, the UC is the only layer for which the solution selection induces a variability larger than the GIGJ estimation error. However, this variability is well below the GIGJ systematic error for the UC, thus concluding that the solution selection has no impact in the geoneutrino signal estimates reported in Table 2.

## 5.3 Model assessment

The GIGJ solution combines local seismic data to constrain the boundaries of the main geophysical discontinuities together with gravity data to overcome the absence of site-specific information.

From a voxel-by-voxel comparison between the GIGJ solution and the prior model, it results that the additional gravity information modifies the shapes of the boundary surfaces and introduces a density variability inside each layer (Figure 7). Regarding the geometry, the higher discrepancies are observed for the TLC and MD for which the mean ± 1σ of the differences between the depth of the layer boundary surfaces are equal to (410 ± 560) m and (110 ± 380) m, with maximum values of 2.2 km and 1.6 km, respectively. Additional gravity data do not change the mean value of crustal density characterizing the prior model, but introduce top-to-bottom vertical gradients with mean density variations of 42.4 kg/m$^3$, 10.8 kg/m$^3$ and 23.6 kg/m$^3$ for the UC, MC and LC, respectively.

A further assessment is performed by comparing the prior and the final model with respect to the available seismic and gravity observations. The prior model fits the DSS profiles with standard deviations of 2.12 km, 2.30 km and 2.51 km for the TMC, TLC and MD, respectively, and fits the gravity dataset with a standard deviation of the residuals $\sigma_g$ = 17.14 mGal. GIGJ preserves the same level of fitting of the seismic profiles, but leads to a gravity interpretation with $\sigma_g$ = 1.14 mGal (Figure 4), fully consistent with the observation noise. Note that a gravity inversion acting only on the geometry of the boundary surfaces without varying the homogeneous densities of the prior model crustal layers, i.e. acting only on the labels $L$ of Eq. (6), is not able to reach the mGal level gravity fitting. From all these considerations, it can be stated that the Bayesian inversion algorithm played a significant role in the final solution by changing the model geometry and density distribution so as to remain coherent with the seismic information, but strongly improving the gravity interpretation.

Thanks to the used local seismic and gravity information, GIGJ is also a significant advance in the modeling of the geophysical structure of the 6° × 4° area centered at JUNO in comparison with the CRUST 1.0 and RRM global models. Although all these models exhibit similar overall crustal thicknesses as a result of the sharp discontinuity given by the Moho, on average the GIGJ model predicts a 30% thinner MC and a 18% thicker LC with respect to global models (Figure 8).

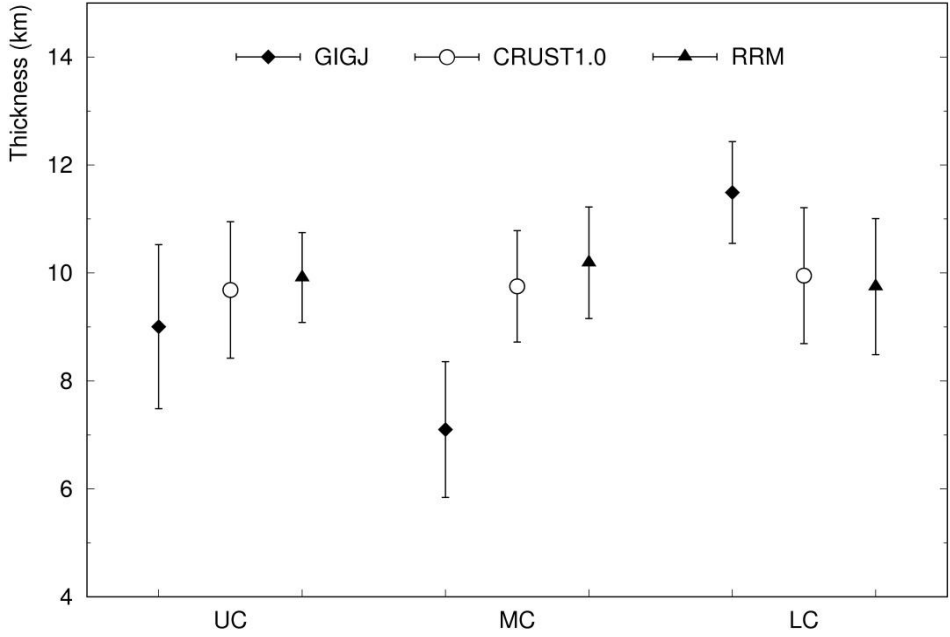

**Figure 8**. Upper Crust (UC), Middle Crust (MC) and Lower Crust (LC) average thicknesses in the 6° × 4° area centered at JUNO, together with their standard deviations, for the GIGJ, CRUST 1.0 and RRM models.

## 6 Expected geoneutrino signal

The three modeled geoneutrino life phases correspond to the production concurrent with beta minus decays along the $^{238}$U and $^{232}$Th decay chains, propagation from the production point to the JUNO location and detection via the inverse beta decay reaction. The GIGJ model was divided into $7 \times 10^7$ cells of 1 km × 1 km × 0.1 km dimensions, each one assigned with crustal layer label, individual density value and unitary U and Th abundances ($a_U$ = 1 μg/g and $a_{Th}$ =1 μg/g).

The expected geoneutrino signal linearly scales with the U and Th mass distributed in the crust and depends on the source-detector distance $r$ by a combined effect of the $1/4\pi r^2$ spherical scaling factor and the average antineutrino survival probability, which oscillations gradually damp for increasing distance from the experimental site. On the basis of the approach and input parameters described in Section 7 of (Strati et al., 2017), we calculated the geoneutrino signal expressed in Terrestrial Neutrino Units (TNU) which correspond to the number of geoneutrino events per $10^{32}$ free target protons per year.

Table 2 summarizes the UC, MC and LC geoneutrino signals ($G_{TOT} = G_U + G_{Th}$) expected at JUNO calculated with GIGJ, the prior model (see Section 4.4), the global CRUST 1.0 and RRM models, always assuming unitary radioisotope abundances.

From the comparison between the signals calculated using the prior and GIGJ models, it is possible to infer that the benefit of using gravity information with the proposed inversion procedure lies in a different repartition of the signal contribution from deep layers (MC and LC) coming from a better understanding the Earth crustal structure below JUNO. Moreover, the adoption of a Bayesian approach for the inversion of the gravimetric problem allowed for estimating the geoneutrino signal uncertainty for each crustal layer by conservatively summing the estimation error and systematic uncertainty on the layer mass (Table 1).

The UC geoneutrino signal is compatible among the GIGJ and global models at 1σ level due to a moderate discrepancy among UC average thicknesses (Figure 8). Nevertheless, the integration of local geophysical input and gravimetric homogenous data according to a Bayesian approach originated a decrease of 77% in the calculated uncertainty on the UC geoneutrino signal. GIGJ predicted a thinner MC and a thicker LC (Figure 8), which implies a reduction (~21%) and an increase (~24%) of the signal with respect to the global models for the MC and LC, respectively. We noted a signal uncertainty reduction of 55% and 78% for MC and LC respectively in comparison with the RRM estimates. The significant improvement for the LC was attributable to the constraint of its bottom surface with consistent local MD depth information used in the a priori model. A refined distinction among the deep crustal layers was a delicate point in the analysis of JUNO geoneutrino data considering that MC and LC are typically characterized by U and Th abundances differing by a factor of ~2.5 and ~13.5 from that of the UC (Huang et al., 2013).

**Table 2**. Total geoneutrino signals $G_{TOT}$ ($G_{TOT} = G_U + G_{Th}$) in TNU assuming unitary uranium and thorium abundances[1].

| | Geoneutrino signal with $a_U = 1$ µg/g and $a_{Th} = 1$ µg/g | | | | | |
|---|---|---|---|---|---|---|
| | GIGJ | | | Prior model | CRUST 1.0 | RRM |
| | $G_U$ (TNU) | $G_{Th}$ (TNU) | $G_{TOT}$ (TNU) | $G_{TOT}$ (TNU) | $G_{TOT}$ (TNU) | $G_{TOT}$ (TNU) |
| UC | 3.25 ± 0.05 | 0.223 ± 0.004 | 3.47 ± 0.05 | 3.47 | 3.56 | 3.72 ± 0.22 |
| MC | 1.59 ± 0.05 | 0.109 ± 0.003 | 1.70 ± 0.05 | 1.47 | 2.09 | 2.20 ± 0.11 |
| LC | 2.03 ± 0.02 | 0.144 ± 0.001 | 2.17 ± 0.02 | 2.39 | 1.73 | 1.77 ± 0.09 |

[1]$G_{TOT}$ values for unitary uranium and thorium abundances are referred to the 6°×4° area centered at the JUNO detector location. $G_{TOT}$ was calculated by adopting the GIGJ, CRUST 1.0 and RRM geophysical crustal models. For the GIGJ model the separate uranium ($G_U$) and thorium ($G_{Th}$) signal components are also reported, where the overall relative uncertainty was obtained by summing the estimation error and systematic relative mass uncertainties (Table 1).

Figure 9 shows the UC, MC and LC maps of the relative contribution to the geoneutrino signal given by each 50 km × 50 km voxel, normalized to the signal produced by each layer. Focusing on the 100 km × 100 km area centered at JUNO, we highlight an evident N-S anisotropy in the UC geoneutrino signal since the two northern voxels originate ~40% of the UC signal. In the perspective of a geophysical and geochemical refinement, this is a relevant information for planning future surveys as done in (Strati et al., 2017).

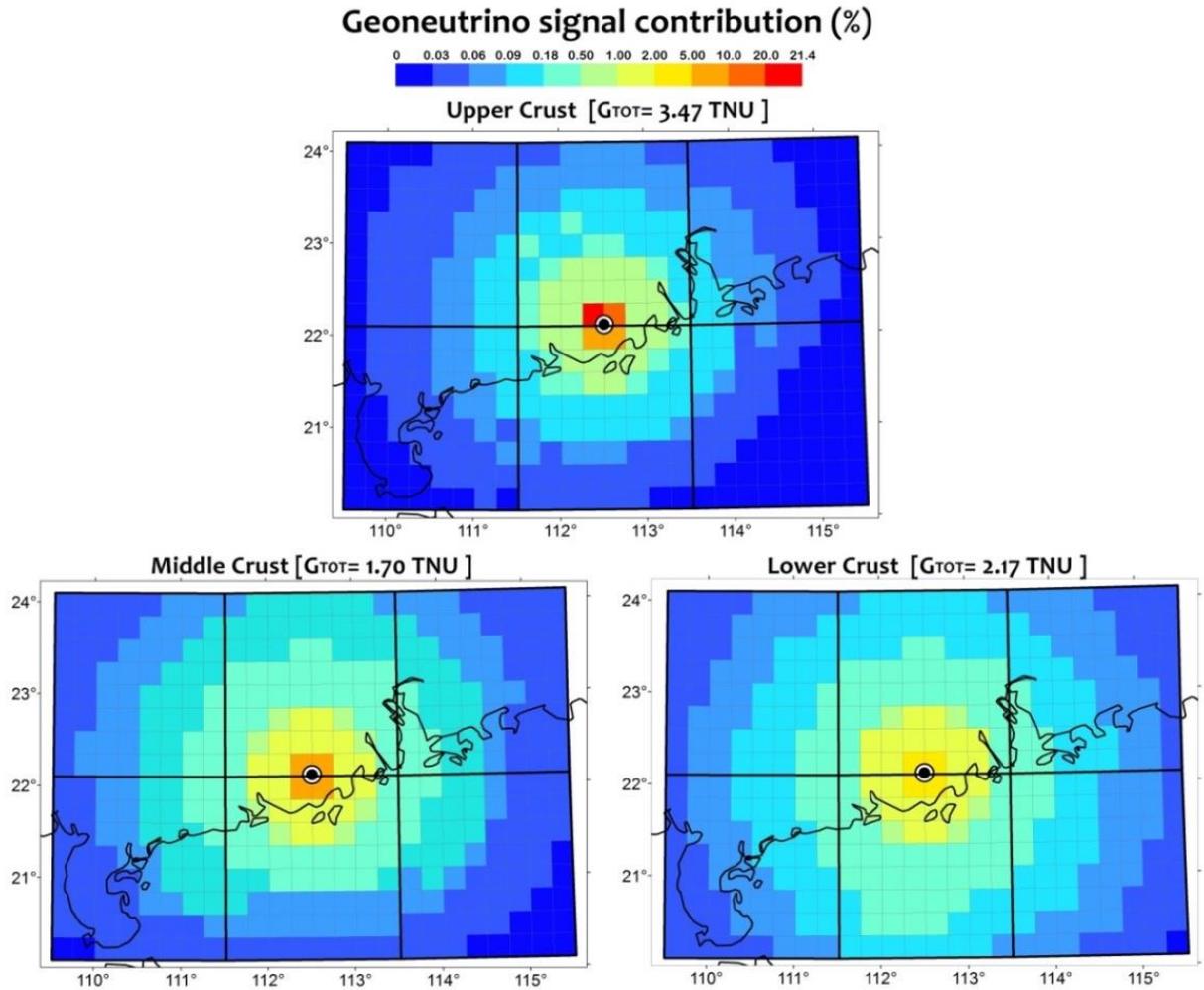

**Figure 9**. Maps of the Upper Crust, Middle Crust and Lower Crust percentage contribution to the total geoneutrino signal ($G_{TOT} = G_U + G_{Th}$) for the 6° × 4° area centered at the JUNO detector location. For each panel, 100% of the signal corresponds to the signal generated by the specific crustal layer.

## 7  Conclusions

A high-resolution 3D voxel-wise crustal model beneath the Guangdong province, south-eastern China, was built by inverting gravimetric data with the aim of predicting the geoneutrino signal at the Jiangmen Underground Neutrino Observatory (JUNO) experiment.

The 6° × 4° area centered at JUNO was studied in the complex tectonic framework of the South China Block. During its history, this block experienced the collision of two Neoproterozoic continental crustal blocks, presently recognized in the Yangtze, forming a stable cratonic area, and in the Cathaysia Block, which comprises the offshore continental margin of the South China Sea. For the first time the heterogeneous information from deep seismic sounding profiles, receiver functions, teleseismic P-wave velocity models and Moho depth maps were integrated into a unique and consistent a-priori geological model used for the gravimetric inversion, that is based on a Bayesian approach and a stochastic optimization. The method was tuned at the algorithm level by supplying a variation range of each depth surface and of each crustal layer density, which was laterally and vertically constrained. The peculiarity of the solution was that it was chosen as the most probable set of labels and densities by taking into account two penalty functions defining the limits of the boundary surfaces and neighborhood rules between different materials.

GIGJ is a 3D numerical model constituted by ~$46\times10^3$ voxels of $50 \times 50 \times 0.1$ km, which is provided in ASCII format at http://www.fe.infn.it/u/radioactivity/GIGJ. GIGJ fitted homogeneously distributed GOCE gravity data with a standard deviation of the residuals of the order of 1 mGal, compatible with the observation accuracy. The solution was the smoothest one in terms of both density distribution and geometrical shape. While global crustal models (e.g. CRUST 1.0) report for UC, MC and LC an equal thickness corresponding to the 33% of the total crustal thickness, GIGJ provides a site-specific subdivision of the crust (Figure 8). GIGJ predicted $M_{UC}$ = [697.8 ± 2.2 (± 9.6)] $10^{16}$ kg, $M_{MC}$ = [585.0 ± 15.8 (± 1.9)] $10^{16}$ kg and $M_{LC}$ = [1001.1 ± 9.3 (± 0.3)] $10^{16}$ kg, where the estimation errors were associated to the solution of the inverse gravimetric problem and the systematic uncertainties (in brackets) were related to the adoption of a fixed sedimentary layer.

Regarding geoneutrino signals prediction, the main outcome of this study was the 77%, 55% and 78% reduction of the UC, MC and LC signal uncertainty. As a consequence of the rearrangement of the crustal layers thicknesses, we predicted a reduction (~21%) and an increase (~24%) of the MC and LC signal respectively, in comparison with the results obtained from global models. This geophysical refinement has strong implications on the predicted local geoneutrino signal, which, once geochemical abundances are taken into account, is foreseen to be about 50% of the total signal (Strati et al. 2015). In perspective, an uncertainty reduction in the expected crustal geoneutrino signal will enhance the potential of JUNO in distinguishing a mantle geoneutrino signal component and in turn testing different bulk silicate compositional models of the Earth.

This study demonstrated that a Bayesian-based gravimetric inversion applied to reliable satellite data rationally integrated with local geological and seismic information provided a coherent picture of the crustal structure at the natural spatial scale required for geoneutrino studies.

## 8    Acknowledgments and Funding

This work was partially founded by National Institute for Nuclear Physics (INFN) through the JUNO experiment, the ITALian RADioactivity project (ITALRAD) and the Theoretical Astroparticle Physics (TAsP) initiative. The co-authors would like to acknowledge the support of the University of Ferrara (FAR-2016 and FIR-2017) and of the Project Agroalimentare Idrointelligente CUP D92I16000030009.

The authors thank the staff of GeoExplorer Impresa Sociale s.r.l. for its support and Kassandra Raptis, Matteo Albèri, Enrico Chiarelli, Carlo Bottardi, Raffaele Tripiccione for their collaboration which made possible the realization of this study. The authors show their gratitude to Eligio Lisi, Livia Ludhova, Paolo Marchetti, William F. McDonough, Fernando Sansò and Oleg Smirnov for useful comments and discussions.

Readers can find the geophysical datasets for the construction of the a priori model together with the GIGJ model athttp://www.fe.infn.it/u/radioactivity/GIGJ.

**Appendix: Geophysical datasets for the construction of the a priori model**

The geophysical surfaces defined for the construction of the a priori model are: the topography/bathymetry, the bottom of sediments, the Top of the Upper Crust (TUC), the Top of the Middle Crust (TMC), the Top of the Lower Crust (TLC), the Moho Discontinuity (MD) and a horizon with a constant depth of 50 km, which is the bottom of the model. In this appendix we provide an enlarged description of the constraints adopted for the definition of the TMC, TLC and MD, which were obtained from published studies, including Deep Seismic Sounding profiles (DSS), Receiver Functions (RF), teleseismic P-wave velocity models and Moho depth maps.

**Deep Seismic Sounding (DSS) profiles**

The P-wave velocities for the different crustal layers were obtained from DSS by digitizing approximately 900 depth-controlling points belonging to 12 seismic profiles (Figure 2 of the main text). The 1σ MD uncertainty was estimated as the picking error from each reference paper. Two quality classes (C1 and C2) for the TLC and TMC uncertainties were defined on the basis of the clarity of the corresponding velocity contour during the digitalization (Table S2).

The LG DSS profile (Zhang & Wang, 2007) is located in the north-eastern part of the study area (Figure 2), crossing the CFB and the SCMB. We adopted the interpretation reported in Figure 3b of (Zhang & Wang, 2007) in which the P-wave events P1 and P4 are respectively identified as the TMC and the TLC corresponding to the velocities reported in Table S1. The same interpretation was also applied to infer the crustal structure for the SC profile (Ji et al., 2017), located in the eastern portion of the CB.

For the south-eastern margin of the SCB, we used the ESP_1_9 and ESP_11_17 (Nissen et al., 1995) profiles (Figure 2). Starting from the velocity-depth models reported in (Nissen et al., 1995), we adopted the velocity contours in Table S1 to define the depths of the TMC, TLC and MD. Along each line, we observed an evident crustal thinning (Table S1) toward the continent-ocean boundary associated with the rifting of the continental crust occurred during the formation of the SCS (Ji et al., 2017).

The OBS2001 DSS profile is located in the transition zone between the CB and the north-eastern part of the of SCS (Figure 2 of the main text) and it is included as input in our analysis adopting the P-wave velocity interfaces imaged in the model reported in (Wang et al., 2006).

As for the other DSS profiles located in the SCS (OBS2006-1, DSRP2002, OOS2004, OBS1993), only a sediment layer (1.5-2.0 km of thickness) and a two-layered structure with a thick UC (11-14 km) on the top of the LC (10-13 km) are interpreted by (Xia et al., 2010) and (Ji et al., 2017). Thus, as the MC is not considered in this model, we defined its boundary as the 6 km/s contour, analogously to the LG DSS profile, which is the closest to the OOS2004 and OBS1993 ones.

Beyond the southern margin of the study area, we used OBHIV that crosses the Xisha Trough, a supposed failed rift with an advancing degree of rifting from west to east (Qiu et al., 2001). Beneath a 1-4 km thick Cenozoic Sedimentary layer, the top of the crust is characterized by a low velocity (Table 1) and the average crustal thickness is 25 km, although in the middle of the section it decreases gradually down to 8 km.

The interpretation of the Bayan_J profile by (Jia et al., 2006) shows an average crustal velocity of 5.7-6.2 km/s to be compared with the 6.2-6.3 km/s which is typical of the SCB. This crustal thinning, as well as the low P-wave velocity values (Table S1) can be linked to the magmatic activities and eruptions that affected this area during the Holocene.

At west of the Hainan Island, we included the onshore-offshore OBH1996-2 DSS profile located in the Yinggehai Basin, by adopting the interpretation reported in (Ji et al., 2017) as crustal structure.

**Teleseismic Receiver Functions (RF)**

Additional punctual constraints for the MD were added on the basis of the analysis of the teleseismic RF. Besides the results obtained from the two stations located inside the study area and close to JUNO (i.e. GZH and SZN, see Figure 2 of the main text), we included also the information of other eight stations located outside (i.e. QIZ, GUL, GYA, CNS, NNC, QZH, WZH, WHN). Following (Tkalčić et al., 2011) we assumed that the most reliable MD were estimated (along with their uncertainties) from GRID SEARCH and H-k methods. Beneath each station we estimated MD as the uncertainty weighted mean of the depth from each method. The $3\sigma$ MD uncertainty accounted for the variability of the different analysis and for the individual uncertainty of each method (Table S2).

**Teleseismic P-wave velocity models**

3D P-wave velocity model from (Sun & Toksöz, 2006) were used to further constrain the MD by digitizing about 90 MD controlling points belonging to the 8 virtual cross sections (C, D, E, F, G I, J and H) reported in Figures 10 and 11 of (Sun & Toksöz, 2006). The $3\sigma$ MD uncertainty was estimated from the final model standard deviation of the travel time corresponding to 0.49 s (Sun & Toksöz, 2006) which, multiplied by 8 km/s (the mean velocity in the lower crust), gave us a MD error of 3.9 km at $1\sigma$ (Table S2).

**Moho depth maps**

The last constraints for the MD came from published models (Hao et al., 2014; He et al., 2013; Xia et al., 2015). Starting from the 1 km interval contours reported in each map, we obtained three regular grids with 10 km × 10 km horizontal resolution.

From the (He et al., 2013) model, we constructed a grid covering the study area and the adjacent areas for a total surface of about 750˙000 km$^2$. The 1σ uncertainty was given as the MD maximum error in our region, obtained from Table 6 of (He et al., 2013).

The Moho depth map of China at scale 1: 5˙000˙000 (Hao et al., 2014) was obtained on the basis of gravity data, while 120 seismic sounding profiles were used as controlling points. The deviations of inversion results of the Moho depth in the investigated region were in the range 1.76-2.24 km. From this map, we obtained a grid including the area that lies within about 650 km from JUNO, and we associated as 1σ uncertainty the maximum MD error.

The last model we considered (Xia et al., 2015) provided a distribution of the MD across the entire CFB. We used this model to create a grid covering an area centered at JUNO with a total surface of about 700˙000 km$^2$. The latter MD model came along with a 1σ uncertainty of 0.5-2 km (Xia et al., 2015).

**Table A1**. Deep Seismic Sounding profiles (DSS) used for the construction of the a priori model. The P-wave velocity contours used for defining the seismic discontinuity (MD = Moho Discontinuity, TLC = Top of the Lower Crust, TMC = Top of the Middle Crust) are reported together with the range of the crustal thickness and the reference paper.

| DSS profile | Range of crustal thickness (km) | MD (km/s) | TLC (km/s) | TMC (km/s) | Reference |
|---|---|---|---|---|---|
| LG | 30 - 34 | 8.0 | 6.7 | 6.2 | (Zhang & Wang, 2007) |
| SC | 29 - 33 | 8.0 | 6.7 | 6.2 | (Ji et al., 2017) |
| ESP_1_9 | 11 - 31 | 8.0 | 6.6 | 6.1 | (Nissen et al., 1995) |
| ESP_11_17 | 16 - 32 | 8.0 | 6.6 | 6.1 | (Nissen et al., 1995) |
| OBS2001 | 16 - 23 | 8.0 | 6.5 | 5.5 | (Wang et al., 2006) |
| OBS2006-1 | 11 - 22 | 8.0 | 6.4 | 6.0 | (Ji et al., 2017) |
| DSRP2002 | 12 - 30 | 8.0 | 6.4 | 6.0 | (Ji et al., 2017) |
| OOS2004 | 24 - 26 | 8.0 | 6.4 | 6.0 | (Xia et al., 2010) |
| OBS1993 | 12 - 26 | 8.0 | 6.4 | 6.0; 5.7 | (Xia et al., 2010) |
| OBHIV | 15 - 25 | 8.0 | 6.4 | 6.1 | (Qiu et al., 2001) |
| Baiyan_J | 25 - 26 | 8.1 | 6.3 | 6.1 | (Jia et al., 2006) |
| OBH1996-2 | 26 - 28 | 8.1 | 6.4 | 6.1; 5.7 | (Ji et al., 2017) |

**Table A2.** Depth uncertainty for the Moho Depth (MD), Top of the Lower Crust (TLC) and Top of the Middle Crust (TMC) adopted in the construction of the a-priori model. The teleseismic P-wave velocity models, the Moho depth maps and the receiver functions are used to parameterize the MD only. The 3σ MD uncertainty is estimated by considering the picking error (1σ) from each reference paper. The uncertainties for the TLC and TMC are subdivided into two quality classes (C1 and C2) based on the clarity of the corresponding velocity contour during the digitalization. The 3σ uncertainties classified as C1 are 1.2 times the 3σ MD uncertainty while the 3σ uncertainties classified as C2 are 1.5 times the 3σ MD uncertainty.

| Type | Input (see Figure 2 of the main text) | MD (km) | TLC (km) | TMC (km) | Reference |
|---|---|---|---|---|---|
| Deep Sounding Seismic profiles | LG | 3.9 | 4.7 (C1) | 4.7 (C1) | (Zhang & Wang, 2007) |
| | SC | 3.9 | 4.7 (C1) | 4.7 (C1) | (Zhang & Wang, 2007) |
| | ESP_1_9 | 3.0 | 4.5 (C2) | 4.5 (C2) | (Nissen et al., 1995) |
| | ESP_11_17 | 3.0 | 4.5 (C2) | 4.5 (C2) | (Nissen et al., 1995) |
| | OBS2001 | 2.4 | 2.9 (C1) | 2.9 (C1) | (Wang et al., 2006) |
| | OBS2006-1 | 4.5 | 6.8 (C2) | 5.4 (C1) | (Xia et al., 2010) |
| | DSRP2002 | 4.5 | 6.8 (C2) | 5.4 (C1) | (Xia et al., 2010) |
| | OOS2004 | 4.5 | 6.8 (C2) | 5.4 (C1) | (Xia et al., 2010) |
| | OBS1993 | 4.5 | 6.8 (C2) | 5.4 (C1) | (Xia et al., 2010) |
| | OBH_IV | 3.6 | 4.3 (C1) | 5.4 (C2) | (Qiu et al., 2001) |
| | Baiyan_J | 2.4 | 2.9 (C1) | 2.9 (C1) | (Jia et al., 2006) |
| | OBH1996-2 | 3.6 | 4.3 (C1) | 4.3 (C1) | (Qiu et al., 2001) |
| Teleseismic P-wave velocity models | Line_C | 11.8 | N/A | N/A | (Sun & Toksöz, 2006) |
| | Line_D | 11.8 | N/A | N/A | (Sun & Toksöz, 2006) |
| | Line_E | 11.8 | N/A | N/A | (Sun & Toksöz, 2006) |
| | Line_G | 11.8 | N/A | N/A | (Sun & Toksöz, 2006) |
| | Line_H | 11.8 | N/A | N/A | (Sun & Toksöz, 2006) |
| | Line_I | 11.8 | N/A | N/A | (Sun & Toksöz, 2006) |
| | Line_J | 11.8 | N/A | N/A | (Sun & Toksöz, 2006) |
| Moho depth maps | Hao_2014_fig3 | 9.0 | N/A | N/A | (Hao et al., 2014) |
| | He_2013 | 9.0 | N/A | N/A | (He et al., 2013) |
| | Xia_2015 | 6.0 | N/A | N/A | (Xia et al., 2015) |
| Receiver Functions | Tel_Tkacic QIZ | 2.1 | N/A | N/A | (Tkalčić et al., 2011) |
| | Tel_Tkacic GZH | 4.8 | N/A | N/A | (Tkalčić et al., 2011) |
| | Tel_Tkacic SZN | 4.8 | N/A | N/A | (Tkalčić et al., 2011) |
| | Tel_Tkacic GUL | 1.8 | N/A | N/A | (Tkalčić et al., 2011) |
| | Tel_Tkacic GYA | 1.5 | N/A | N/A | (Tkalčić et al., 2011) |
| | Tel_Tkacic CNS | 2.5 | N/A | N/A | (Tkalčić et al., 2011) |
| | Tel_Tkacic NNC | 5.1 | N/A | N/A | (Tkalčić et al., 2011) |
| | Tel_Tkacic QZH | 1.5 | N/A | N/A | (Tkalčić et al., 2011) |
| | Tel_Tkacic WZH | 8.5 | N/A | N/A | (Tkalčić et al., 2011) |
| | Tel_Tkacic WHN | 3.3 | N/A | N/A | (Tkalčić et al., 2011) |